\documentclass[twocolumn,a4paper,10pt,showpacs,prl,aps,superscriptaddress,amsmath,amssymb,amsfonts,floatfix]{revtex4-1}
\usepackage[utf8x]{inputenc}
\usepackage{graphicx, array}
\usepackage{amsmath}
\usepackage{rotating}
\usepackage[caption=false]{subfig}

\begin{document}
\title{Moir{\'e} patterns as a probe of interplanar interactions: graphene on h-BN}

\author{M. M. van Wijk}
\affiliation {Radboud University Nijmegen, Institute for Molecules and Materials, Heyendaalseweg 135, 6525 AJ Nijmegen, The Netherlands}
\author{A. Schuring}
\affiliation {Radboud University Nijmegen, Institute for Molecules and Materials, Heyendaalseweg 135, 6525 AJ Nijmegen, The Netherlands}
\author{M. I. Katsnelson}
\affiliation {Radboud University Nijmegen, Institute for Molecules and Materials, Heyendaalseweg 135, 6525 AJ Nijmegen, The Netherlands}
\author{A. Fasolino}
\affiliation {Radboud University Nijmegen, Institute for Molecules and Materials, Heyendaalseweg 135, 6525 AJ Nijmegen, The Netherlands}

\begin{abstract}
By atomistic modeling of moir{\'e} patterns of graphene on a substrate with a small lattice mismatch, we find  qualitatively  different strain distributions  for small and large misorientation angles, corresponding to the commensurate-incommensurate transition recently observed in graphene on hexagonal BN. We find that the ratio of C-N and C-B interactions is the main parameter determining the different bond lengths in the center and edges of the  moir{\'e} pattern. Agreement with experimental data is obtained only by assuming that the C-B interactions are at least twice weaker than the C-N interactions. 
The correspondence between the strain distribution in the nanoscale moir{\'e} pattern and the potential energy surface at the atomic scale found in our calculations, makes the moir{\'e} pattern a tool to study details of dispersive forces in van der Waals heterostructures. 
\end{abstract}


\maketitle

After the discovery of graphene, many other layered materials have been identified which can be exfoliated to form single or few-layer systems~\cite{novoselov2005two}. Layers of different materials can be combined in precise sequences to form what have been called van der Waals heterostructures~\cite{geim2013heterostructures}. The study of these new hybrid materials is emerging as a strong research area.

The superposition of periodic layered structures, with either slightly different lattice constants or different orientations, creates moir{\'e} patterns~\cite{yankowitz2012emergence, ponomarenko2013cloning, dean2013hofstadter, hunt2013massive, woods2014commensurate}. These patterns can yield a wealth of information about the lattice constant mismatch, strain and imperfections of the surface~
\cite{hattab2012interplay,gai1996application,carter1995dislocations,pushpa2002stars,rockett1991energetics,chen2012molecular}.
The moir{\'e} patterns imply a change of the interatomic distances that can affect properties that are important both for applications and for fundamental physics such as the quantum mechanics of electrons in quasi-periodic potentials~\cite{yankowitz2012emergence, ponomarenko2013cloning, dean2013hofstadter, hunt2013massive}.

In recent years hexagonal boron nitride (h-BN) has become a standard substrate for graphene growth due to its flat surface without dangling bonds, the hexagonal lattice with a lattice constant only 1.8 \% larger than that of graphene and the fact that h-BN is an insulator~\cite{dean2010boron}. These properties have led to the realization of the first field effect transistor~\cite{britnell2012field}. 
The difference in lattice constant leads to the appearances of moir{\'e} patterns, which can be observed experimentally~\cite{tang2013precisely,xue2011scanning,yang2013epitaxial}.

Usually, moir{\'e} structures are considered from a purely geometrical point of view for the superposition of two rigid lattices where the length $L$ of the moir{\'e} patterns is found to depend on the angle $\theta$ and the lattice mismatch between the two layers as 
\begin{equation}
 L = \frac{p}{\sqrt{1+p^2-2p \cos(\theta)}}a,
\end{equation}
where $p$ is the ratio between lattice constants and $a$ the lattice constant of the substrate~\cite{hermann2012periodic}.
Strain due to the lattice mismatch and/or rotations have been considered in a continuum approach to study the modification of the electronic structures in tight binding calculations~\cite{mucha2013heterostructures,bistritzer2011moire,sanjose2014electronic,sanjose2014spontaneous,cosma2014} and the pseudo-magnetic fields resulting from out-of-plane displacements~\cite{neekamal2014moire, neekamal2014stress}. Full atomic relaxation to minimal energy configurations is however necessary to make a detailed comparison to experimental structural information as obtained by  scanning probe microscopy~\cite{woods2014commensurate}. At the same time, we will show that this procedure allows to get quantitative information on the interplanar interactions. It is well known that dispersive forces are beyond the standard local density functional and generalized gradient corrections~\cite{spanu2009nature}. Several attempts have been made to calculate dispersive interactions between graphene and h-BN using more rigorous approaches~\cite{sachs2011adhesion, bokdam2014band,jung2014abinitio}.

Recently, evidence for an incommensurate-commensurate transition in graphene on h-BN at a critical rotation angle has been found by scanning probe and Raman spectroscopy~\cite{woods2014commensurate}. The authors  examined moir{\'e} patterns with periodicity ranging between $L=8$~nm and $L=14$~nm ($\theta \sim 1.5^\circ-\sim 0^\circ$) 
and found a sudden change of the strain distribution in the moir{\'e} pattern at $L\sim 10$~nm. 
At large angles (small moir{\'e} pattern) the Young modulus distribution displays a sinusoidal behavior whereas at small angles (large moir{\'e} pattern) it presents sharp peaks on a constant baseline. This change of behavior was attributed to the evolution from an incommensurate structure with continuous small adjustment of the graphene lattice to locally commensurate domains separated by narrow domain walls~\cite{woods2014commensurate}. These two situations, found for large and small angle respectively, originate from two competing energy terms. The dispersive (van der Waals) interaction with the substrate favors stretching of the graphene to adapt to the underlying h-BN whereas the interactions within the layer favor the graphene equilibrium bond length. 

In this work, we present a fully atomistic model to compute both in-plane and out-of-plane atomic displacements and the distribution of strain in graphene on a substrate. In view of the large moir{\'e} periodicity at small angles, one needs to consider very large supercells which are not only much beyond the possibility of ab-initio calculations but may be also very demanding for classical atomistic approaches based on empirical potentials. In particular, imposing periodic boundary conditions for a specific value of the strain in layers rotated by a very small angle can easily lead to cells made of millions of atoms. Therefore, in the following we will consider the specific strain of the graphene/h-BN system only for $\theta=0$, where the commensurate situation should occur, and for a rather large angle $\theta$. Typically we need to deal with tens of thousands atoms per layer. 

An atomistic approach allows to examine the distortions and establish a comparison to experiment. It turns out that the behavior of in-plane and out-of-plane distortions is very sensitive to the interplanar interactions. In a sense, the moir{\'e} patterns take the role of a magnifying glass which projects the interatomic interactions at their larger length scale.  

We study, by energy minimization, the adaptation of a graphene layer to a substrate with the same hexagonal structure but a different lattice constant, representing h-BN as discussed below. We choose a rotated and unrotated case to examine the commensurate-incommensurate transition reported in Ref.~\cite{woods2014commensurate}. 

The graphene atoms interact through the REBO potential~\cite{brenner2002second} as implemented in the molecular dynamics code LAMMPS~\cite{lammps}. For this potential the equilibrium bond length of graphene is 1.3978 \AA. The h-BN substrate is kept rigid, mimicking a bulk substrate. No empirical potential for graphene on h-BN is available. The interplanar potential energy calculated ab-initio~\cite{sachs2011adhesion} is however qualitatively similar to the one of graphite, with a minimum of about 20 meV/atom  at a distance which is much larger than the one for covalent bonding and similar to the interplanar distance of graphite. Therefore, we begin by  modeling h-BN as stretched graphene. Pair potentials like  Lennard-Jones underestimate the corrugation of the interplanar potential energy surface~\cite{reguz2012potential,neekamal2014stress}. For this reason, we describe the interaction between graphene and h-BN by a registry-dependent potential for graphene~\cite{RDP0}, which is scaled to the lattice constant of h-BN.  We minimize the total potential energy by relaxing the graphene layer by means of FIRE~\cite{bitzek2006fire}, a damped dynamics algorithm. 
We model the unrotated case ($\theta=0$) by  56 $\times$ 56 unit cells of graphene on 55 $\times$ 55 unit cells of h-BN, resulting in a  1.8 \% mismatch in lattice constant.

Constructing a coincident lattice for two rotated graphene layers can be done by rotating one of them from $\mathbf{r}=n\mathbf{a_1}+m\mathbf{a_2}$ to $\mathbf{t}=m\mathbf{a_1}+n\mathbf{a_2}$ with $n,m$ integer, which fixes the angle $\theta$ and the number of atoms $N$ in the cell~\cite{shallcross2010electronic, savini2011bending}. The smallest cell that can be obtained is the one with ($n,m$)=(2,1). We then scale the lattice constant of the bottom layer to the lattice constant of h-BN and repeat this cell $55 \times 55$ times while we do not scale the top layer and repeat it $56 \times 56$ times. In this way, we obtain a supercell with $N=86254$ and $\theta\approx38^\circ$. 

\begin{figure}[htbp]
 \centering
 \subfloat[]{\includegraphics[width=0.5\linewidth]{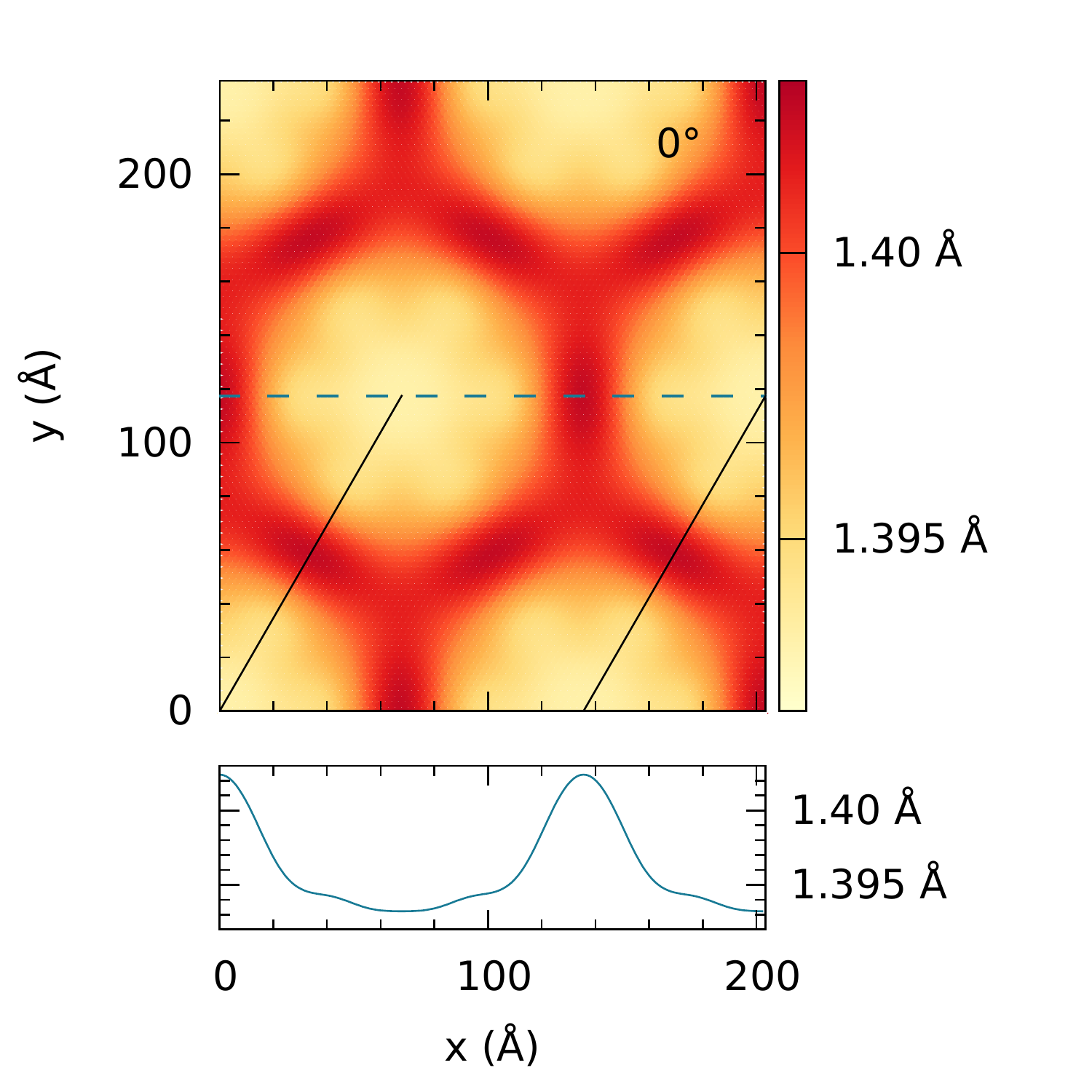}}
 \subfloat[]{\includegraphics[width=0.5\linewidth]{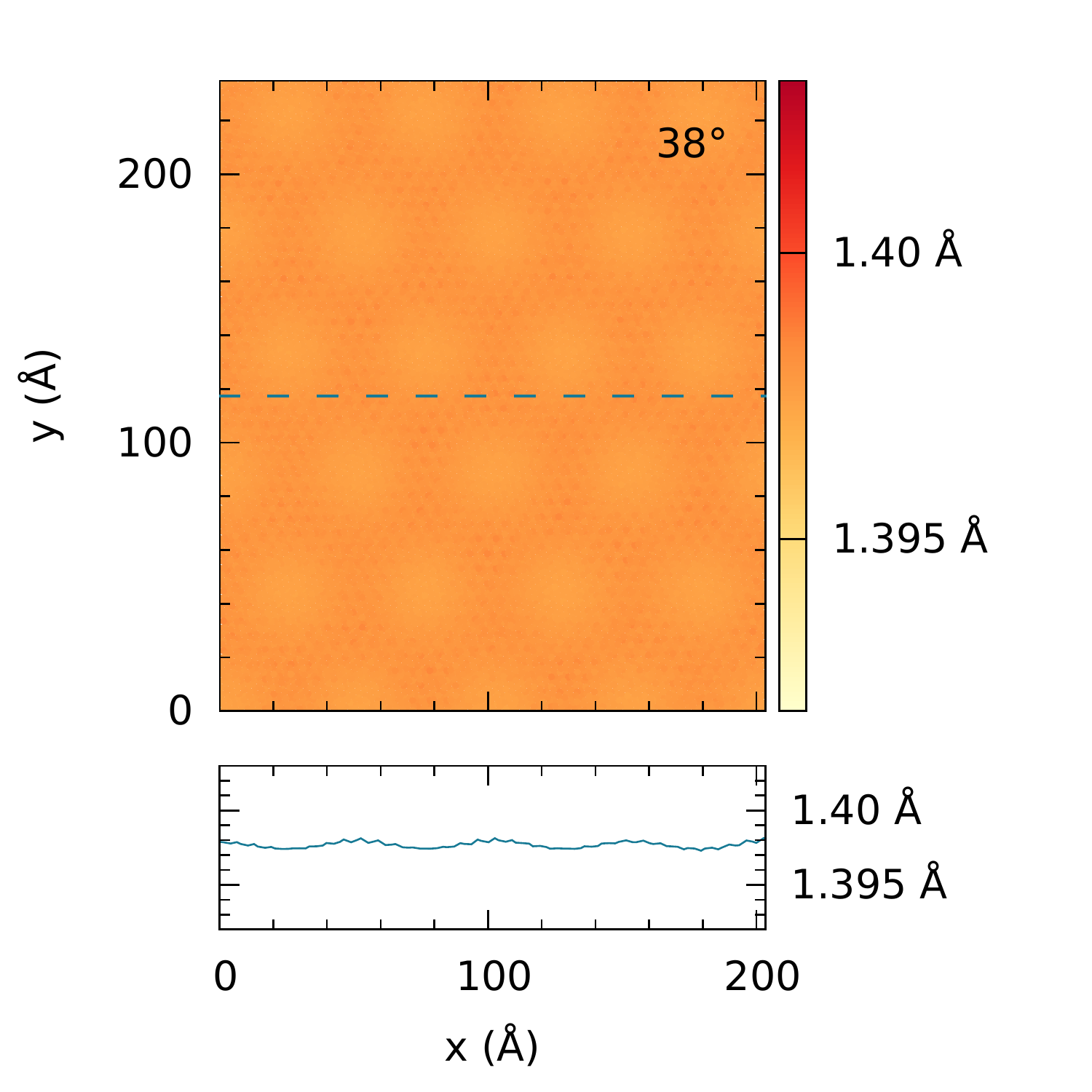}}
 \caption{(color online) (top panels) Color coded distribution of bond lengths in a graphene layer on a rigid h-BN substrate; (bottom panels) bond lengths along the horizontal dashed line shown in the top figure. (a) $\theta= 0^\circ$, $N=12322$. The supercell of side $L=135.6$ \AA~is indicated by the solid black line. (b) $\theta=38^\circ$, $N=86254$. We show only part of the supercell with a side of $358.7$ \AA$>> L=6.5$\AA.}
 \label{fig:comincom}
\end{figure}

Very different results are obtained for $\theta=0^\circ$ and $\theta=38^\circ$. We show the distribution of bond lengths for these angles in Fig.~\ref{fig:comincom}. While for $\theta=0^\circ$ clear differences in bond length are visible throughout the moir{\'e} pattern, the bond lengths for the large angle are much more homogeneous. At first glance these results seem in agreement with the experiments~\cite{woods2014commensurate} but actually there is a very important difference. While our simulations show a smaller lattice constant in the center of a moir{\'e} hexagon and a larger one at the edges, the opposite is found in the experiment. 

\begin{figure}[htbp]
 \centering
\includegraphics[width=0.9\linewidth]{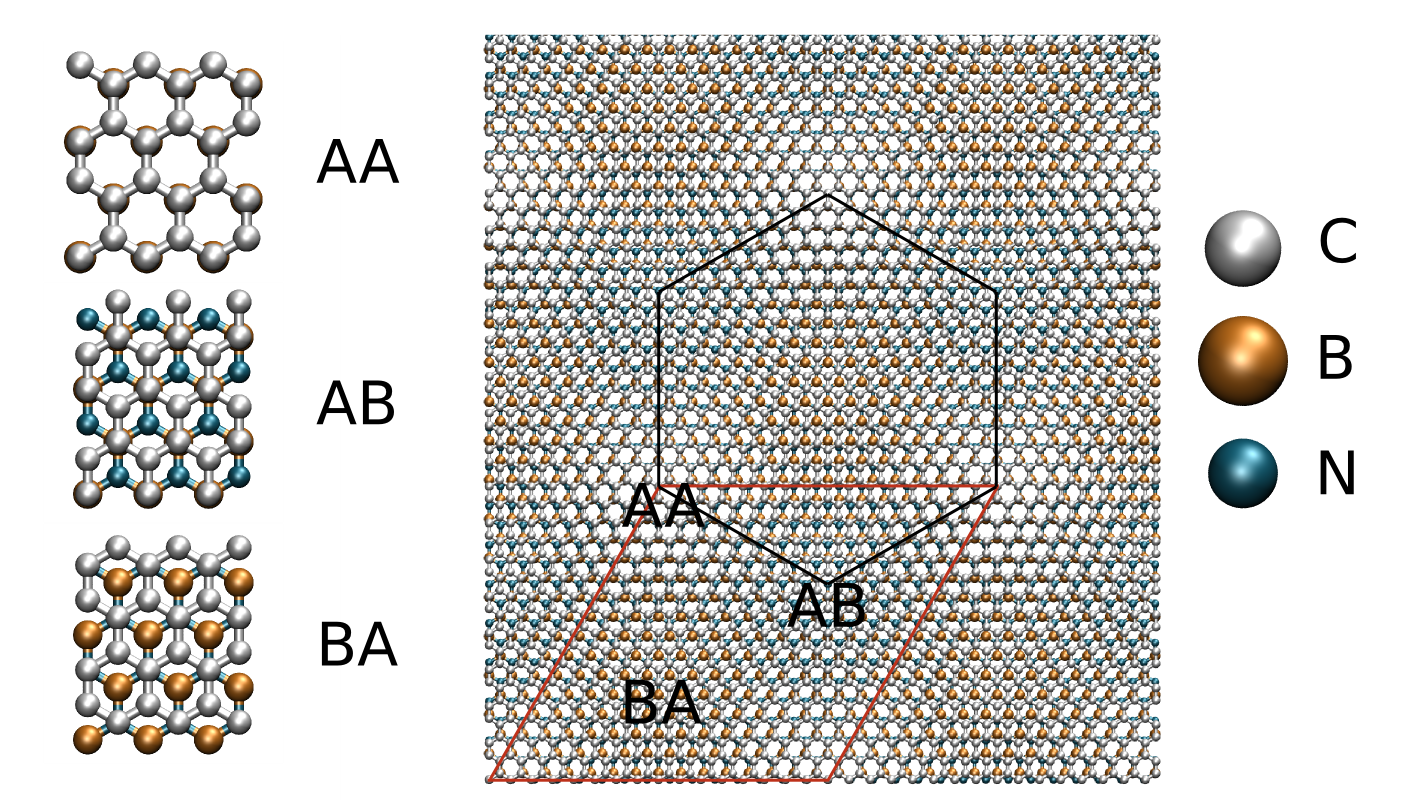}
 \caption{(color online) Different stackings and moir{\'e} pattern for graphene on h-BN. The difference in lattice constant is exaggerated for clarity. The supercell is shown in red and the moir{\'e} pattern in black. 
}
 \label{fig:moirehbn}
\end{figure}

The driving force for the commensurate-incommensurate transition should be the tendency to minimize the interlayer energy  by adopting the lattice constant of the substrate at the expense of creation of domain walls. Out-of-plane distortions result from this process, but they cannot lead to a commensurate-incommensurate transition since a larger lattice constant of graphene in the central area of the moir{\'e} pattern is required for commensurability. In this sense, the experimental data is intuitively clear and it is unexpected that our model, albeit simplified, gives such a qualitative difference. The hexagonal lattice with two atoms per cell is not a Bravais lattice and this turns out to be crucial as we explain next. 
Ab-initio calculations~\cite{sachs2011adhesion,bokdam2014band} show that the interactions between graphene and h-BN are dominated by the C-N interaction. The configuration where a N atom sits in the center of a graphene hexagon (AB stacking, see Fig.~\ref{fig:moirehbn}) was found to be the most energetically favorable. At the same time, the configuration where a B atom sits in the center of a graphene hexagon (BA stacking) was only slightly better than the one with all atoms sitting on top of other atoms (AA stacking). To model this situation, we  vary the strength of the C-B interaction by scaling the potential to $s=$50~\%, 30~\%, 10~\% and 0~\% of the C-N interaction. In this way, we go over from a hexagonal lattice on a hexagonal lattice ($s=100$~\%) to a hexagonal lattice on a triangular lattice ($s=0$~\%). 

\begin{figure*}[ht!]
\centering
\includegraphics[width=\linewidth]{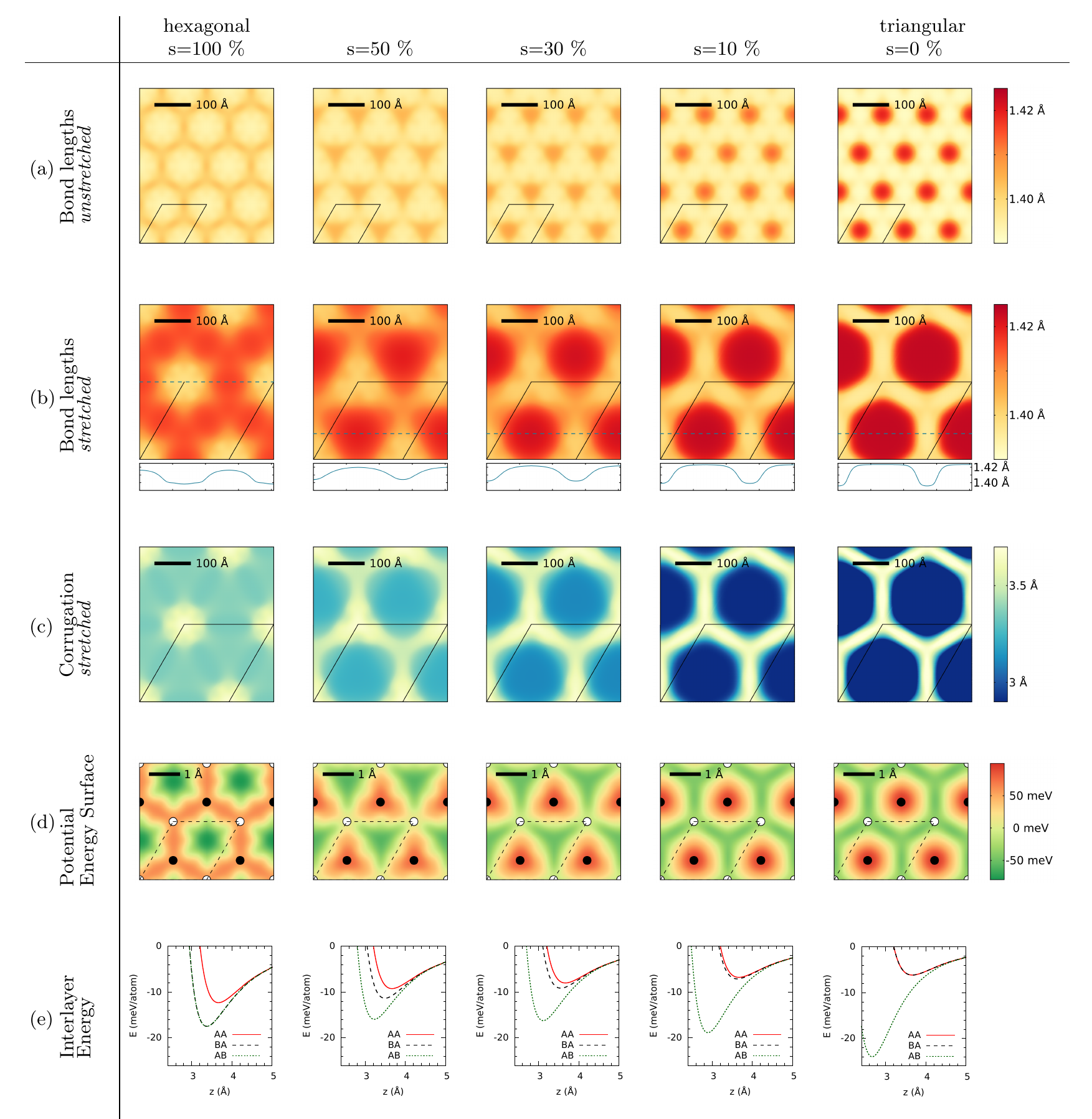}
\caption{(color online) Several properties as a function of the interaction strength of C-B with respect to C-N for $\theta=0^\circ$. The supercell is shown with a continuous black line. (a) Distribution of bond lengths in an \textit{unstretched} graphene layer on a rigid h-BN substrate. (b) Distribution of bond lengths in an \textit{stretched} graphene layer on a rigid h-BN substrate and bond lengths along the horizontal dashed line shown in the top figure. (c) Distance of the \textit{stretched} graphene layer to the substrate. (d) Potential energy surface at a constant height of 3 \AA. The unit cell is shown with a dashed line, N atoms in black and B atoms in white. Note the difference in scale. (e) Interlayer energy as a function of the distance between the layers $z$ for different stackings.  }
\label{fig:interact}
\end{figure*}

Furthermore, only the {\it relative} difference of the lattice constants in the center and edges of the moir{\'e} pattern has been measured~\cite{woods2014commensurate}. Therefore we consider also a graphene layer stretched globally by 0.9\% which we call \textit{stretched} in Fig.~\ref{fig:interact}. We show below that this global stretching has only a quantitative effect on the size ratio of the central region to the edges. The asymmetry in C-N and C-B interactions instead is crucial to reproduce the observed strain distribution.

Figs.~\ref{fig:interact}a,b,c show that the strain distribution and out-of-plane displacements depend dramatically on the ratio $s$ of the C-B/C-N interactions. The size and hexagonal shape of the moir{\'e} pattern do not change, but the distribution of bond lengths (Fig.~\ref{fig:interact}a,b) and the distance to the substrate (Fig.~\ref{fig:interact}c) strongly depend on the interaction ratio. For the hexagonal substrate ($s=100 \%$), the strained part is located at the edges of the moir{\'e} pattern, whereas for the triangular lattice ($s=0 \%$) the center is stretched.  The source of this difference is clarified by Fig.~\ref{fig:moirehbn}. The graphene is stretched to adapt its lattice constant to the one of h-BN in the areas with the most favorable stacking. If the AB is the most favorable while both AA and BA are unfavorable ($s=0 \%$), AB is the center of a hexagonal moir{\'e} pattern and adaptation to h-BN will take place there. If instead AB and BA are equally favorable ($s=100 \%$), AA is at the center of a hexagon and the stretching will occur at the edges of the moir{\'e} pattern. 
In Fig.~\ref{fig:interact} we show the gradual changes from one to the other situation for intermediate values of $s$. 
A global stretching of graphene (Fig.~\ref{fig:interact}b) only  makes the areas with larger bond lengths wider. The out-of-plane displacements (corrugation) shown in (Fig.~\ref{fig:interact}c) follow qualitatively the same trend as the in-plane displacements.
For $s=$10~\%, 30~\%, 50~\% we find that the amplitude of the out-of-plane displacements is 0.79~\AA, 0.56~\AA~and 0.44~\AA~respectively, against $\sim$ 0.5~\AA~experimentally~\cite{yang2013epitaxial}.
 
The changes of the strain distribution, described above as a function of the C-B/C-N interaction ratio, are mirrored at the atomic scale of a single unit cell in the potential energy surfaces (PESs) shown in Fig.~\ref{fig:interact}d. The color code gives the energy of an atom moving over the h-BN unit cell at a constant height of 3~\AA.
While changing $s$ the maxima (green) and minima (red) of the PESs are interchanged in the same way as the strain distribution, making the moir{\'e} patterns a magnified image of the interplanar interactions at the atomic scale. For completeness, in Fig.~\ref{fig:interact}e we show the interlayer energy of two rigid layers for different stackings. The shape of the strain distribution in Fig.~\ref{fig:interact}a,b is similar to experiment~\cite{woods2014commensurate} for $s \sim 50$\% and lower, implying that the C-B interaction is two to three times weaker than the C-N interaction. Weaker C-B interactions also yield out-of-plane distortions in better agreement with experimental data~\cite{yang2013epitaxial}. 

In summary, we suggest that the strain distribution and out-of-plane displacement in moir{\'e} patterns give direct information on the interplanar  interactions in van der Waals heterostructures. For the case of graphene on h-BN, we demonstrated different adjustment to the substrate for large and small moir{\'e} patterns, supporting the commensurate-incommensurate transition found experimentally~\cite{woods2014commensurate}. We showed that the distributions of bond lengths in the pattern are strongly dependent on the ratio between carbon-boron and carbon-nitrogen interactions. Comparison to experiment implies that the carbon-boron interaction is two to three times weaker than the carbon-nitrogen interaction. 

\noindent
$ {\bf Acknowledgements} $. This work is part of the research program of the Foundation for Fundamental Research on Matter (FOM), which is part of the Netherlands Organisation for Scientific Research (NWO). The research leading to these results has received funding from the European Union Seventh Framework Programme under grant agreement n°604391 Graphene Flagship. We thank Kostya Novoselov and Yury Gornostyrev for useful 
discussions.

\end{document}